\documentclass[aps]{revtex4}

\def\oversim#1#2{\lower0.2ex\vbox{\baselineskip0pt\lineskip0pt
  \lineskiplimit0pt\ialign{$#1\hfil##\hfil$\crcr#2\crcr\sim\crcr}}}
\newcommand{\simgt}{\lower.5ex\hbox{$\; \buildrel > \over \sim \;$}}
\newcommand{\simlt}{\lower.5ex\hbox{$\; \buildrel < \over \sim \;$}}
\newcommand{\nablara}{\raisebox{10pt}{$\rightarrow$}\hspace{-1.em}
                   \raisebox{-2pt}{$\nabla$}}
\newcommand{\nablala}{\raisebox{10pt}{$\leftarrow$}\hspace{-1.em}
                   \raisebox{-2pt}{$\nabla$}}
\newcommand{\nablalr}{\raisebox{10pt}{$\leftrightarrow$}\hspace{-1.em}
                   \raisebox{-2pt}{$\nabla$}}

\begin{document}

\title{Covariant Wigner Function Approach to the Boltzmann \\
Equation for Mixing Fermions in Curved Spacetime}

\author{Kazuhiro Yamamoto}

\bigskip
\affiliation{
Department of Physics, Hiroshima University,
Higashi-Hiroshima 739-8526,~Japan\\
}



\begin{abstract}
Based on the covariant Wigner function approach
we derive the quantum Boltzmann equation for fermions 
with flavor mixing in general curved spacetime.
This work gives a rigorous theoretical framework
to investigate the flavor oscillation phenomena 
taking the gravitational effect into account.
It is shown that the Boltzmann equation of the 
lowest order of the expansion with respect to $\hbar$ 
reproduces the previous result which was derived in 
the relativistic limit on the Minkowski background 
spacetime.
It is demonstrated that the familiar formula for 
the vacuum neutrino oscillation can be obtained 
by solving the Boltzmann equation. Higher order
effects of the $\hbar$-expansion is also briefly discussed.
\end{abstract}

\maketitle

\def\M{{M}}
\section{Introduction}

The Boltzmann equation, which describes non-equilibrium systems 
of many particles, is one of the most fundamental to investigate 
physical processes in the early universe. 
For example, the statistics of the angular anisotropies in the 
cosmic microwave background is predicted by solving the Boltzmann 
equation for photons in curved spacetime (see. e.g., \cite{Hu}).
The decoupling of heavy particles from thermal medium in the expanding 
universe, which is known as the generalized Lee-Weinberg problem,
is investigated with the use of the Boltzmann equation
\cite{Bernstein,KolbTurner}. 
The Boltzmann equation for leptogenesis is a recent topics of such early 
universe physics \cite{BP,Endoh}. 
Because field theory is more fundamental to describe particle processes,
it is useful to find the method to derive the Boltzmann equation 
from the field theory.  
There are two different approaches to derive the Boltzmann equation from 
the field theory. One is the method with the use of the density matrix
and the other is the Wigner function approach, which we will adopt in 
the present paper.
Thus the Boltzmann equation is regarded as an effective theory of the 
field theory in a many-particle limit. 
(see e.g. the text book \cite{Groot}.)

On the other hand, recently, the flavor mixing of the neutrinos 
has been experimentally established by several groups \cite{SK}.
These results strongly suggest that the neutrinos have non-zero mass
and that their interaction basis is not the mass eigenstate.
The propagation of particles with mass mixing might not be trivial 
in general curved spacetime. 
Actually the gravity effect on the neutrino oscillation has been 
investigated by several authors (\cite{NOG,Bha,Wudka,Konno} and references therein). 
However, these works focus on how the quantum-mechanical phase of 
the mixing is affected due to the gravity. We might claim that the 
Boltzmann equation is needed to investigate collective 
phenomena of many-particle systems.
The Boltzmann equation of the neutrinos has been investigated 
by several authors (see \cite{SR,SP,Yamada} and references therein). 
Sigl and Raffelt developed a general formula of the Boltzmann 
equation for the neutrinos with flavor mixing using the density 
matrix approach \cite{SR}. 
It has been shown that the same result can be reproduced with 
the Wigner function approach \cite{SP,Yamada}. 
The Boltzmann equation of the neutrinos is very important to 
understand the early universe. For example, Dolgov et~al. have
claimed that the flavor equilibrium can be achieved due to the 
flavor mixing effect before the neutrino decoupling, which might 
lead a significant constraint on the lepton number asymmetry of 
the background neutrinos by combining a constraint from the 
successful primordial nucleosynthesis \cite{DR,Dolgov}.

The Boltzmann equation of the neutrinos, in the previous works,
have been derived assuming the Minkowski background spacetime.
The gravity effect on the Boltzmann equation is usually taken into 
account by replacing the Liouville operator in the Minkowski spacetime
with that in curved spacetime. The validity of this procedure might 
be worth being checked. 
The primary aim of the present paper is to derive the Boltzmann 
equation for fermions with flavor mixing in general curved 
spacetime in a rigorous manner starting from the Dirac equation 
on curved spacetime. 
For that purpose, we take the covariant Wigner function 
approach.
It is known that a simple Wigner function approach encounters 
the problem of the covariance of the general coordinate 
transformation. To avoid this problem, several authors considered 
the covariant Wigner function. Winter defined the
covariant Wigner function by introducing the covariant geodesic 
distance \cite{Winter}, while Calzetta, Habib and Hu defined
the covariant Wigner function using the Riemann normal coordinate 
\cite{CHH}.
Fonarev developed a very elegant framework for the covariant
Wigner function using the tangent space. He checked that 
these three methods yield the locally same equation \cite{FonarevSP,FonarevJMP}. 
Here we adopt the framework by Fonarev to apply it to the fermions with 
flavor mixing.

This paper is organized as follows: In section 2 we introduce
the covariant Wigner function for mixing fermions, and derive 
its equation of motion. 
The equation of motion can be regarded as the Boltzmann 
equation for mixing fermions in curved spacetime. 
Some details of the derivation are described in the Appendix.
In section 3, we show that the same equation is reproduced 
at the lowest order of the expansion with respect to $\hbar$ 
in the ultra-relativistic limit, which is a useful test of 
the previous result using the density matrix formalism.  
In section 4, we demonstrate that the familiar formula of 
the transition probability of the neutrino oscillation
is obtained by solving the Boltzmann equation.
Higher order effects of the $\hbar$-expansion 
on the Boltzmann equation is discussed in section 5. 
Section 6 is devoted to summary and conclusions.

In the present paper we use the unit in which the light velocity $c$ equals $1$.
We follow the convention of $(+---)$ for the metric $g_{\mu\nu}$.
We use $\alpha,\beta,\mu,\nu,\cdots$ to denote the index of coordinate,
$\Gamma^\alpha_{\mu\nu}$ is the Christoffel symbol,
and the Riemann tensor is defined
$R^{\alpha}{}_{\beta\mu\nu}
  =\partial_\mu\Gamma^\alpha_{\beta\nu}
  -\partial_\nu\Gamma^\alpha_{\beta\mu}
  +\Gamma^\alpha_{\mu\sigma}\Gamma^\sigma_{\nu\beta}
  -\Gamma^\alpha_{\nu\sigma}\Gamma^\sigma_{\mu\beta}$.
In this paper, the superscript, $A,B,C\cdots$ are used to denote the flavor index, 
while spinor index is omitted. 

\section{Covariant Wigner Function}
\def\hnabla{{\hat \nabla}}
\def\cnabla{{\check \nabla}}
\def\tr{{\rm tr}}
\def\hatD{{D}}

In this section we introduce the covariant Wigner function 
following Fonarev \cite{FonarevSP}. 
The covariance for the general coordinate transformation is manifest 
due to the definition with the use of the tangent space \cite{FonarevJMP}. 
In order to introduce the Wigner function, it is instructive to 
start from considering the energy momentum operator for the 
fermion fields $\psi^A$ \cite{BD}:
\begin{eqnarray}
  T_{\mu\nu}={i\over 2}\tr[\bar\psi^B\gamma_{(\mu}\nabla_{\nu)}\psi^A-
                        \nabla_{(\mu}\bar\psi^B\gamma_{\nu)}\psi^A],
\end{eqnarray}
where the subscript $(\mu\cdots\nu)$ denotes the symmetrization with respect to $\mu$ 
and $\nu$,  $A$ and $B$ denote the flavor index, and $\tr$ should be
taken for the spinor and the flavor indices. The gamma matrix satisfies
\begin{equation}
  \gamma^\mu\gamma^\nu+\gamma^\nu\gamma^\mu=2g^{\mu\nu}.
\end{equation}
Note that the above expression can be rephrased 
\begin{eqnarray}
  T_{\mu\nu}=-\tr \biggl[
  \gamma_{(\mu}\psi^A\Bigl(-{i\hbar \over 2}\nablalr_{\nu)}
  \Bigr)\bar\psi^B\biggr],
\end{eqnarray}
where $\nablalr_{\nu}=\nablara_\nu-\nablala_\nu$. 
Thus, in general, we consider an operator $\hat A$ which 
can be written as
\begin{eqnarray}
  {\hat A} = -\tr \biggl[\Gamma \psi^A 
  a\Bigl( -{i\hbar \over 2}\nablalr_\nu \Bigr)\bar\psi^B
  \biggr],
\end{eqnarray}
where $\Gamma$ consists of $\gamma$-matrix and $a$ is a function 
of $-{i\hbar}\nablalr/2$. 
Then we find that ${\hat A}$ can be written as
\begin{eqnarray}
  {\hat A}=-\tr {\Gamma \over (\pi\hbar)^4}\int d^4p {1\over \sqrt{-g(x)}}
  \int d^4 y \sqrt{-g(x)} e^{-2iy^\alpha p_\alpha/\hbar} 
   a(p_\nu)
  \Psi^A(x,-y)\bar\Psi^B(x,y),
\end{eqnarray}
where 
\begin{eqnarray}
\Psi^A(x,-y)&=&\Bigl(1-y^\alpha\nabla_\alpha
  +{1\over 2!}y^\alpha y^\beta\nabla_\alpha\nabla_\beta-\cdots\Bigr)\psi^A(x),
\label{Psip}
\\
\bar \Psi^B(x,y)&=&\Bigl(1+y^\alpha\nabla_\alpha
  +{1\over 2!}y^\alpha y^\beta\nabla_\alpha\nabla_\beta+\cdots\Bigr)
  \bar\psi^B(x).
\label{Psim}
\end{eqnarray}
The expectation value of $\hat A$ is written 
\begin{eqnarray}
  \langle {\hat A}\rangle=\tr \int {d^4p \over  \sqrt{-g(x)}} \Gamma a(p_\nu) 
  N^{AB}(x,p),
\end{eqnarray}
where $N^{AB}(x,p)$ is the covariant Wigner function defined by 
\begin{eqnarray}
   N^{AB}(x,p)={-1\over (\pi \hbar)^4}\int d^4y \sqrt{-g(x)} 
  e^{-2iy^\alpha p_\alpha /\hbar}
  \langle \Psi^A(x,-y)\bar\Psi^B(x,y)\rangle.
\label{defN}
\end{eqnarray}
For example the expectation value of the energy momentum tensor is
\begin{eqnarray}
  \langle T_{\mu\nu}\rangle=\tr \int {d^4p \over  \sqrt{-g(x)}} 
  \gamma_{(\mu}p_{\nu)} N^{AB}(x,p).
\end{eqnarray}
In general curved spacetime, $\langle T_{\mu\nu}\rangle$ diverges.
However, how to regularize the divergence is not explicitly 
specified here (see also \cite{FonarevJMP}).

Equation of motion of $N^{AB}(x,p)$ is the (generalized) 
Boltzmann equation, which we derive from the Dirac 
equation for $\psi^A(x)$. 
In the present paper, we consider the Dirac equation:
\begin{eqnarray}
  \sum_B[i\hbar \gamma^\mu \nabla_\mu \delta^{AB} -M^{AB}(x)
  -\hbar \gamma^\mu J_\mu^{AB}(x)]
  \psi^B(x)=0,
\label{DiracE}
\end{eqnarray}
where $J_\mu^{AB}(x)$ denotes an effective potential which arises 
from interaction with background particles \cite{RR}. 
Here we adopt the 
above expression motivated by the interaction of the neutrinos, which
are mediated by the vector bosons. 

We briefly summarized the derivation of equation of motion for $N^{AB}(x,p)$ 
in Appendix for being self-contained, though the derivation is similar to 
that in the reference \cite{FonarevJMP}. Equation of motion for 
$N^{AB}(x,p)$ is formally obtained by the expansion with respect 
to the power index of $\hbar$. The equation of 
motion up to the order of $\hbar^2$ is 
\begin{eqnarray}
  &&\gamma^\alpha \biggl(p_\alpha+{i\hbar\over 2}\hatD_\alpha \biggr)N^{AB}(x,p)
  -\sum_C\biggl(M^{AC}(x)+\hbar \gamma^\mu J_{\mu}^{AC}(x)
 \nonumber
\\
  &&\hspace{2cm}
 -{i\hbar\over 2}\nabla_\alpha \Bigl(M^{AC}(x)+\hbar \gamma^\mu J_{\mu}^{AC}(x)\Bigr)
  {\partial \over \partial 
  p_\alpha}
 +{\hbar^2\over 8}  \nabla_{\nu_1}\nabla_{\nu_2}M^{AC}(x)
  {\partial\over \partial p_{\nu_1}}{\partial\over \partial p_{\nu_2}}
  \biggr)N^{CB}(x,p)
\nonumber
\\
  &&=-\gamma^\alpha\biggl(
  {\hbar^2\over 16}{\partial N^{AB}(x,p)\over \partial p_\nu}
  R_{\nu\alpha\mu\rho}\sigma^{\mu\rho}
\nonumber
\\
  &&\hspace{2cm}
  -{\hbar^2\over 24}p^\mu R_{\nu_1\alpha\nu_2\mu}
   {\partial\over \partial p_{\nu_1}}{\partial\over \partial p_{\nu_2}}N^{AB}(x,p)
  -{7\hbar^2\over 48} R_{\alpha\mu}
  {\partial\over \partial p_{\mu}}N^{AB}(x,p)
\biggr),
\nonumber
\\
\end{eqnarray}
where we defined 
\begin{eqnarray}
  D_\alpha=\nabla_\alpha+\Gamma^\beta_{\alpha\nu}p_\beta
  {\partial\over\partial p_\nu}
\label{defD}
\end{eqnarray}
and $\sigma^{\mu\rho}=[\gamma^\mu,\gamma^\rho]/4.$
We first consider the equation up to the order of $\hbar$
in the next section.
The effect of the higher order terms in proportion to $\hbar^2$ 
is discussed in section 5.

\section{Equation of $O(\hbar)$}
In this section we focus on the equation of order $\hbar$:
\begin{eqnarray}
  &&\gamma^\alpha \biggl(p_\alpha+{i\hbar\over 2}\hatD_\alpha \biggr)
  N(x,p)
  -\biggl(M(x)+\hbar \gamma^\alpha J_{\alpha}(x)
  -{i\hbar\over 2}\nabla_\alpha M(x) {\partial \over \partial 
  p_\alpha}
  \biggr)N(x,p)=0,
\label{eqnab}
\end{eqnarray}
where we omit the flavor index of $N^{AB}(x,p)$, for simplicity.
Note that we omit the spinor index too. Following the procedure 
in \cite{SP}, we separate the left handed chirality component from the
right handed component.  Introducing the left- and right-
handed chirality projection operators
\begin{eqnarray}
&&P_{L}={1-\gamma^5\over 2},
\\
&&P_{R}={1+\gamma^5\over 2},
\end{eqnarray}
respectively, where
\begin{eqnarray}
\gamma^5=-{1\over 4!}\sqrt{-g(x)}\epsilon_{\alpha\beta\mu\nu}
  \gamma^\alpha\gamma^\beta\gamma^\mu\gamma^\nu,
\end{eqnarray}
we define
\begin{eqnarray}
N_{RL}(x,p)&=&{P_R}N(x,p)P_R,
\\
N_{L}(x,p)&=&{P_L}N(x,p)P_R,
\\
N_{R}(x,p)&=&{P_R}N(x,p)P_L,
\\
N_{LR}(x,p)&=&{P_L}N(x,p)P_L.
\end{eqnarray}
In the present paper, in general, we consider $J_\mu$ written in the form:
\begin{eqnarray}
  J_\mu(x)=P_L J_{L\mu}(x)+P_R J_{R\mu}(x),
\end{eqnarray}
where $J_{L\mu}(x)$ and $J_{R\mu}(x)$ are the vector quantities.  
Projecting $P_L$ and $P_R$ from left and right on equation (\ref{eqnab}), 
respectively, we have
\begin{eqnarray}
  &&\gamma^\alpha\biggl(p_\alpha+{i\hbar\over 2}\hatD_\alpha 
  -\hbar J_{R\alpha}(x)\biggr)N_{RL}
  -\biggl(M(x)-{i\hbar\over 2}\nabla_\alpha M(x) 
  {\partial \over \partial  p_\alpha}
  \biggr)N_L(x,p)=0.
\label{eqNab}
\end{eqnarray}
In a similar way, 
projecting $P_R$ and $P_R$ from left and right on equation (\ref{eqnab}),
we have 
\begin{eqnarray}
  &&\gamma^\alpha \biggl(p_\alpha+{i\hbar\over 2}\hatD_\alpha 
  -\hbar J_{L\alpha}(x)\biggr)N_{L}
  -\biggl(M(x)-{i\hbar\over 2}\nabla_\alpha M(x) 
  {\partial \over \partial  p_\alpha}
  \biggr)N_{RL}(x,p)=0.
\label{eqNac}
\end{eqnarray}
Combining (\ref{eqNab}) and (\ref{eqNac}), we eliminate
$N_{RL}(x,p)$, and obtain the equation for $N_{L}$ 
up to the order of $\hbar$
\begin{eqnarray}
  &&\biggl(p^\alpha p_\alpha-M^2(x)+i\hbar p^\alpha \hatD_\alpha +
  {i\hbar\over 2}(\nabla_\alpha M^2(x)){\partial \over \partial  p_\alpha}
 -\hbar p_\alpha J_{L\beta}(x)\gamma^\alpha \gamma^\beta\
\nonumber
\\
  &&\hspace{2cm}  -\hbar M(x)J_{R\alpha}(x)M^{-1}(x)p_\beta 
  \gamma^\alpha \gamma^\beta\
  -i\hbar(\nabla_\alpha M(x))M^{-1}(x)p_\beta \gamma^\alpha \gamma^\beta
  \biggr)N_L(x,p)
  =0,
\nonumber
\\
\label{eqNad}
\end{eqnarray}
where $M^{-1}(x)$ is the inverse of the mass matrix $M(x)$. Similar to the way
to derive (\ref{eqNad}), equations for $N_R(x,p)$
can be derived, which have the same 
expression as (\ref{eqNad}) but with replacing $R(L)$ by $L(R)$.

Now we focus on equation (\ref{eqNad}). Let us consider a
solution of the form:
\begin{equation}
  N_L(x,p)=F_\mu(x,p)P_L\gamma^\mu.
\label{Fmu}
\end{equation}
Note that $N_L(x,p)$ has the spinor index, but
$F_\mu(x,p)$ does not have it. Inserting (\ref{Fmu}) into
(\ref{eqNad}), we finally have 
\begin{eqnarray}
  &&\biggl(p^\alpha p_\alpha-M^2(x)+i\hbar p^\alpha \hatD_\alpha +
  {i\hbar\over 2}(\nabla_\alpha M^2(x)){\partial \over \partial  p_\alpha}
  \biggr)F_\nu(x,p) 
\nonumber
\\
  &&\hspace{1cm}-K_{\alpha}^{~\alpha}F_\nu(x,p)
  +K_{\mu\nu}F^\mu(x,p)-K_{\nu\mu}F^\mu(x,p)
  -i\epsilon^{\alpha\beta\mu}_{~~~~~\nu}K_{\alpha\beta}F_{\mu}(x,p)
  =0,
\label{eqNae}
\end{eqnarray}
where we defined
\begin{eqnarray}
  K_{\alpha\beta}=\hbar p_\alpha J_{L\beta}
  +\hbar M(x)J_{R\alpha}(x)M^{-1}(x)p_\beta 
  +i\hbar(\nabla_\alpha M(x))M^{-1}(x)p_\beta.
\end{eqnarray}

\section{Application to Neutrinos}
In this section, we apply the result of the previous section
to the neutrinos. 
Setting $\nabla_\alpha M(x)=0$ and $J_{R\alpha}(x)=0$, 
equation (\ref{eqNae}) reduces to
\begin{eqnarray}
  &&\Bigl(p^\alpha p_\alpha-M^2+i\hbar p^\alpha \hatD_\alpha 
  \Bigr)F_\nu
 -\hbar \Bigl(p_\alpha J_{L}^\alpha F_\nu -p_\alpha J_{L\nu} F^\alpha
  +p_{\nu}J_{L\alpha}F^\alpha+i\epsilon^{\alpha\beta\mu}_{~~~~\nu}
  p_\alpha J_{L\beta}F_\mu\Bigr)=0.
\nonumber
\\
\label{eqNaf}
\end{eqnarray}
This is consistent with the result in the reference \cite{SP}, 
in which the effect of the gravity is not taken into account. 
As we will show below, in the ultra-relativistic regime, 
we may set
\begin{eqnarray}
  F_\nu=p_\nu F,
\label{FnuF}
\end{eqnarray}
where $F=F(x,p)$ is a scalar function. Substituting (\ref{FnuF})
into $(\ref{eqNaf})$, we have
\begin{eqnarray}
  p_\nu(p^\alpha p_\alpha-M^2+i\hbar p^\alpha D_\alpha-2\hbar p^\alpha 
  J_{L\alpha})F+\hbar p^\alpha p_\alpha J_{L\nu}F=0.
\label{eqNag}
\end{eqnarray}
The last term of the left hand side of the above equation
can be neglected in the ultra-relativistic limit of $(p^0)^2\gg 
O(M^2)$. In this limit, equation of $F$ is 
\begin{eqnarray}
  (p^\alpha p_\alpha-M^2-2\hbar p^\alpha J_{L\alpha}
  +i\hbar p^\alpha D_\alpha)F=0.
\label{eqNah}
\end{eqnarray}
Note that $F$ has the flavor index and that $F$ is hermite. 
Taking the dagger of (\ref{eqNah}),
\begin{eqnarray}
  F(p^\alpha p_\alpha-M^2-2\hbar p^\alpha J_{L\alpha})
  -i\hbar p^\alpha D_\alpha F=0.
\label{eqNai}
\end{eqnarray}
Here we assumed $J_{L\alpha}^\dagger=J_{L\alpha}$. 
{}From (\ref{eqNah}) and (\ref{eqNai}) we have
\begin{eqnarray}
  &&\{p^\alpha p_\alpha-M^2-2\hbar p^\alpha J_{L\alpha}, F\}=0,
\label{onshell}
\\
  &&i\hbar p^\alpha D_\alpha F=-{1\over 2}[F,M^2+2\hbar J_{L\alpha} p^\alpha],
\label{Boltzmann}
\end{eqnarray}
where $\{\cdots,\cdots\}$ and $[\cdots,\cdots]$ denote
the anticommutator and commutator, respectively. 
Equation (\ref{onshell}) is the constraint equation
to describe the on-shell condition or the dispersion relation,  
while (\ref{Boltzmann}) is the Boltzmann equation. 
This result is consistent with the previous result 
in the flat spacetime limit \cite{SR,SP,Yamada}.
The gravity effect of this Boltzmann equation arises
through the differential operator $p^\alpha D_\alpha$,
with which we can write 
\begin{eqnarray}
  p^\alpha D_\alpha F(x,p)&=&p^\alpha\left({\partial \over \partial x^\alpha}+
  \Gamma^\beta_{\alpha\nu}p_\beta {\partial \over \partial p_\nu}\right) F(x,p)
\nonumber
 \\
  &=& {d\over d\lambda} F(x(\lambda),p(\lambda)),
\label{affinepara}
\end{eqnarray}
where $\lambda$ is the affine parameter. This suggests that 
the affine parameter will be fundamental to describe the
oscillation in general curved spacetime.
{}From now on, we use the unit $\hbar=1$ for convenience,
excepting the case we write $\hbar$ explicitly to avoid 
confusion.

\def\aalpha{\xi}
\def\bbeta{\zeta}

Here we demonstrate that the familiar formula of 
the neutrino oscillation probability
is obtained by solving the Boltzmann equation.
For simplicity, we consider the two flavor system with the mass 
matrix 
\begin{eqnarray}
  &&M^2=\left(\begin{array}{cc}
\cos\theta & -\sin\theta\\
\sin\theta & \cos\theta
\end{array}\right)
\left(\begin{array}{cc}
m_1^2 & 0\\
0 & m_2^2\\
\end{array}\right)
\left(\begin{array}{cc}
\cos\theta & \sin\theta\\
-\sin\theta & \cos\theta\\
\end{array}\right)
\end{eqnarray}
and the source term $J_{L\alpha}$
\begin{eqnarray}
&&J_{L\alpha}=\left(\begin{array}{cc}
  J_{L\alpha}^{ee}& 0\\
  0& J_{L\alpha}^{\mu\mu}
\end{array}\right).
\end{eqnarray}
In the ultra-relativistic limit, we assume that $F$ might be written
\begin{eqnarray}
  F=
 \left(\begin{array}{cc}
 f_{ee}     & f_{e\mu}\\
 f_{e\mu}^* & f_{\mu\mu}
\end{array}\right).
\end{eqnarray}
Writing the off-diagonal component $f_{e\mu}$ as
\begin{eqnarray}
  f_{e\mu}=Re[f_{e\mu}]+iIm[f_{e\mu}],
\end{eqnarray}
(\ref{Boltzmann}) yields
\begin{eqnarray}
  && p^\alpha D_\alpha f_{ee}=2\bbeta Im[f_{e\mu}],
\label{bolA}
\\
  && p^\alpha D_\alpha f_{\mu\mu}=-2\bbeta Im[f_{e\mu}],
\label{bolB}
\\
  && p^\alpha D_\alpha Im[f_{e\mu}]=-\bbeta (f_{ee}-f_{\mu\mu})
  +2\aalpha Re[f_{e\mu}],
\label{bolC}
\\
  && p^\alpha D_\alpha Re[f_{e\mu}]=-2\aalpha Im[f_{e\mu}],
\label{bolD}
\end{eqnarray}
where we defined
\begin{eqnarray}
&&\aalpha={1\over 4}(m_2^2-m_1^2)\cos2\theta 
  -{1\over 2}(J_{L\alpha}^{ee}-J_{L\alpha}^{\mu\mu})p^\alpha,
\\
&&\bbeta={1\over 4}(m_2^2-m_1^2)\sin2\theta.
\end{eqnarray}

We next demonstrate that the above equations reproduce the 
well-known formula of the neutrino oscillation 
on the flat spacetime background. 
We consider one dimensional stationary flux of neutrinos along 
the $z$ axis, where the momentum of the relativistic neutrinos 
has the component $p^\alpha=(p,0,0,p)$. 
In this case, combining equations 
(\ref{bolA}-\ref{bolD}), we have (see also \cite{SP})
\begin{eqnarray}
  &&{\partial^3  \over \partial z^3} (f_{ee}-f_{\mu\mu})
    -{1\over \xi}{\partial  \xi \over \partial z}
    {\partial^2  \over \partial z^2} (f_{ee}-f_{\mu\mu})
    +4{\xi^2+\zeta^2\over p^2}
    {\partial  \over \partial z} (f_{ee}-f_{\mu\mu})
    -{4\over \xi}{\partial  \xi \over \partial z}
     {\zeta^2\over p^2}(f_{ee}-f_{\mu\mu})=0
\nonumber
\\
\label{bolE}
\end{eqnarray}
and
\begin{eqnarray}
  && {\partial  \over \partial z}( f_{ee}+f_{\mu\mu})=0.
\label{bolF}
\end{eqnarray}
Equation (\ref{bolF}) yields $f_{ee}+f_{\mu\mu}=$constant,
which means the total number of the neutrinos conserves.
We must specify initial condition to find a solution. 
We adopt the initial condition so that the electron 
neutrinos are emitted at $z=0$, namely,
$f_{ee}(z=0)=f_{ee}(0)$, $f_{\mu\mu}(z=0)=0$, and $f_{e\mu}(z=0)=0$. 
In the vacuum case $J_{L\alpha}=0$, integration of (\ref{bolE}) yields 
\begin{eqnarray}
  &&{\partial^2  \over \partial z^2} f_{\mu\mu}
  +\omega^2 f_{\mu\mu}=2{\zeta^2\over p^2}f_{ee}(0),
\label{bolG}
\end{eqnarray}
where $\omega=(m_2^2-m_1^2)/2p$.
Further integration gives the solution 
\begin{eqnarray}
  {f_{\mu\mu}(z)\over f_{ee}(0) }=\sin^2 2\theta \sin^2 
  \left({(m_2^2-m_1^2)z \over 4p}\right),
\label{bolH}
\end{eqnarray}
which gives the familiar oscillation probability from 
the electron neutrino to the mu neutrino. 

The oscillation probability is given by considering the phase 
evolution of the mass eigenstates in the familiar approach. 
Namely, assuming a wave function that can be decomposed into a 
coherent superposition of the mass eigenstate, the oscillation 
probability is given by evaluating the transition amplitude
from an initial state to other state in flavor space.  
On the other hand, our approach is based on the Boltzmann
equation. The equivalence of the both approaches
is not clearly proved here, however, the above 
demonstrates the usefulness of the Boltzmann approach. 
Below we brielfy discuss the effect of the gravity on the phase shift 
by comparing with previous works 
as a test of our framework.

Consider a trajectory of a particle $x^\alpha(\lambda)$ parameterized 
by the affine parameter $\lambda$.
Because we are considering the ultra-relativistic limit, 
we assume that the trajectory is a null geodesic.
Then, from equation (\ref{affinepara}), the Boltzmann 
equation in the vacuum is written 
\begin{eqnarray}
  &&{d^3  \over d\lambda^3} (f_{ee}-f_{\mu\mu})
    +\left({m_2^2-m_1^2\over 2}\right)^2 
    {d  \over d\lambda} (f_{ee}-f_{\mu\mu})
    =0
\label{bolM}
\end{eqnarray}
and
\begin{eqnarray}
  && {d \over d\lambda}(f_{ee}+f_{\mu\mu})=0,
\label{bolN}
\end{eqnarray}
where $f_{ee}$ and $f_{\mu\mu}$ are understood as the 
functions along the trajectory
$f_{ee}=f_{ee}(x(\lambda),p(\lambda))$ and
$f_{\mu\mu}=f_{\mu\mu}(x(\lambda),p(\lambda))$, respectively.
Along the trajectory we have the formal solution 
\begin{eqnarray}
  && f_{ee}-f_{\mu\mu}=C_1 + C_2\sin\left({m_2^2-m_1^2\over 2}\lambda(x,p)+\delta\right)
\label{bolY}
\end{eqnarray}
and
\begin{eqnarray}
  && f_{ee}+f_{\mu\mu}=C_3,
\label{bolZ}
\end{eqnarray}
where $C_1$, $C_2$, $C_3$ and $\delta$ are the constants 
which should be determined by initial conditions. Thus 
the phase of the general solution is proportional to the 
affine parameter, which may contain the gravitationally 
induced contribution along the geodesic. This result can
be consistent with \cite{Wudka}, in which 
a resolution for the controversy in the references 
\cite{NOG,Bha} is presented (see also \cite{Konno}). 
However, our argument assumes that the wave function 
of the different mass eigenstates overlap along the null 
geodesics. 
To incorporate the on-shell condition of the different mass 
eigenstates explicitly in our framework, 
we need to introduce the distribution function using 
the basis of the mass eigenstate \cite{SP}.
We will revisit this problem beyond the ultra-relativistic
limit as a future work. 


\section{Higher order effect}

Here we briefly discuss the higher order effect of
the $\hbar$-expansion of the Boltzmann equation.
As we have shown in section 2, the new terms
of the order of $\hbar^2$ are
\begin{eqnarray}
  &&{{i\hbar^2 \over 2}
  \gamma^\mu \nabla_\alpha J_{\mu}(x) {\partial\over \partial p_\alpha}}N(p,x)
\label{highone}
\end{eqnarray}
and 
\begin{eqnarray}
  {\hbar^2\gamma^\alpha R_\alpha}[N(p,x)],
\label{hightwo}
\end{eqnarray}
where we defined
\begin{eqnarray}
{R_\alpha}[N(p,x)]&=&{1\over 64}
 {\partial N(p,x)\over \partial p^\nu}R_{\nu\alpha\mu\rho}[\gamma^\mu,\gamma^\rho]
\nonumber
\\
&&-{1\over 24}p^\mu R_{\nu_1\alpha\nu_2\mu}
{\partial\over \partial p_{\nu_1}}{\partial\over \partial p_{\nu_2}}N(p,x)
-{7\over 48}R_{\alpha\mu}{\partial\over \partial p_{\mu}}N(p,x).
\end{eqnarray}
Similar to section 3, the left- and right- handed components can be 
separated. Equations corresponding to (\ref{eqNab}) and (\ref{eqNac}) 
are
\begin{eqnarray}
  &&\gamma^\alpha\left(p_\alpha+{i\hbar\over 2}D_\alpha\right)N_{RL}+
  {\hbar^2R_\alpha}[N_{RL}]-MN_L=0,
\\
  &&\gamma^\alpha \left(p_\alpha+{i\hbar\over 2}D_\alpha-\hbar J_{L\alpha}
+{{i\hbar^2 \over 2}
  \nabla_\mu J_{L\alpha}(x) {\partial\over \partial p_\mu}}\right)N_{L}
 +{\hbar^2\gamma^\alpha R_\alpha} [N_{L}]
  -MN_{RL}=0,
\end{eqnarray}
where we consider the case $\nabla_\alpha M(x)=0$ and $J_{R\alpha}(x)=0$. 
Combining these equations, 
\begin{eqnarray}
  &&\left( p^\alpha p_\alpha-M^2 +i\hbar p^\alpha D_\alpha -\hbar \gamma^\alpha \gamma^\beta
  p_\alpha J_{L\beta}\right) N_L
\nonumber\\
  &&+{i\hbar^2\over 2} \gamma^\alpha\gamma^\beta\left(
  p_\alpha (\nabla_\mu J_{L\beta}) {\partial N_L\over \partial p_\mu}
 -D_\alpha (J_{L\beta}N_L)\right)
\nonumber\\
  &&-{\hbar^2\over 4}\gamma^\alpha \gamma^\beta D_\alpha D_\beta N_L
  +\hbar^2 \left( \gamma^\alpha \gamma^\beta p_\alpha R_{\beta}[N_L]
  +\gamma^\alpha R_\alpha[\gamma^\beta p_\beta N_L]
  \right)=0.
\end{eqnarray}
Note that the quantum effect due to spacetime curvature appears as 
the correction of the order of $\hbar^2$, i.e., expression (\ref{hightwo}).  
According to the 
analogy with the equation of the order of $\hbar$, the effect of 
the spacetime-curvature alters the constraint equation (on-shell 
condition) \cite{FonarevJMP}, 
while the variation of the effective potential, (\ref{highone}), 
affects the dynamical equation.
However, the correction of the higher order terms are small in general cases
because the correction is of order of $\hbar/pL$, where $L$ is 
a typical length scale of spacetime curvature or the 
gradient of the effective potential, and $p$ is the momentum 
of a particle. Thus $\hbar/pL$ is the ratio of the
de Broglie wave length to the curvature length scale, 
which is generally small, except for e.g., 
very early epoch of the universe.

\section{Summary and Conclusions}

In this paper we have derived the quantum kinetic equation
for the fermions with flavor mixing in curved spacetime.
This derivation is based on the covariant Wigner function 
approach developed by Fonarev \cite{FonarevJMP}. 
The result presents a rigorous theoretical framework
to investigate flavor oscillation phenomena taking the 
gravitational effect in general curved-spacetime 
into account. The formula is expressed in terms of the
expansion with respect to the power index of $\hbar$.
The new terms of the order of $\hbar^2$ are the quantum
effect due to the gravity, which alter the on-shell 
condition. 
At present the physical consequence of the correction on the 
on-shell condition is not clear (cf. see \cite{FonarevPLB,Borner}).
The equation of the order of $\hbar$ is consistent with
the previous results. We have shown that the equation of 
the order of $\hbar$ reduces to the previous result 
\cite{SR,SP,Yamada}, with simply replacing the Liouville 
operator in curved spacetime to that in the Minkowski spacetime. 
It is also demonstrated that the familiar formula for the vacuum 
neutrino oscillation probability can be obtained by solving the 
Boltzmann equation. 
Then it is shown that our approach gives the consistent results 
which relies on the phase evolution. 
However, these results assume the ultra-relativistic limit. 
We have also derived the general Boltzmann equation 
which does not assume the ultra-relativistic limit in 
section 3. Analysis of the equation will clarify the effects 
of the finite mass on the neutrino's propagation in general
curved spacetime.  We will revisit this problem in future work.

\acknowledgements{
Some parts of this work were done during the stay at Max-Planck Institute
for Astrophysics (MPA), which was supported by a fellowship for Japan
Scholar and Researcher Abroad from Ministry of Education, Sports, Science
and Culture of Japan. The author thanks Prof. S. D. M. White and the
people at MPA for their hospitality during his stay. He also thanks Y. 
Kojima, T. Morozumi, T. Murata, and K. Endoh for useful conversations 
related to the topics in the present paper. The author is grateful to
Prof. D. V. Ahluwalia for useful comments on the manuscript, which helped
improve it.}

\vspace{1cm}
\begin{appendix}

\section{Mathematical Formulas}
\def\tnabla{{\tilde \nabla}}
\def\atnabla{{\larrow{\tilde \nabla}}}
\def\calM{{\cal M}}
\def\calK{{\cal K}}
\def\calR{{\cal R}}
\def\calA{{\cal A}}

Here we derive equation for the covariant Wigner function $N^{AB}$ 
defined in (\ref{defN}).
The derivation is similar to that in the reference \cite{FonarevJMP},
excepting our generalization of the mass term and the flavor index,
however, we write this Appendix for being self-contained.
It is useful to define the derivative operator:
\begin{equation}
  \hnabla_\alpha=\nabla_\alpha-\Gamma^{\beta}_{\alpha\gamma}y^\gamma
  {\partial\over \partial y^\beta},
\end{equation}
then we find 
\begin{equation}
  \hnabla_\alpha y^\beta=0.
\end{equation}
Equations (\ref{Psip}) and (\ref{Psim}) are, respectively, written as
\begin{eqnarray}
   &&\Psi^A(x,-y)=\exp(-y^\alpha\hnabla_\alpha)\psi^A(x)
\end{eqnarray}
and
\begin{eqnarray}
  &&\bar\Psi^B(x,y)=\exp(y^\alpha\hnabla_\alpha)\bar\psi^B(x).
\end{eqnarray}

For arbitrary operators $\hat A$ and $\hat B$, the identities
\begin{eqnarray}
  &&[\hat A,e^{\hat B}]={-}\sum_{n=1}^\infty{1\over n!}
  [\hat B,[\cdots,[\hat B,\hat A]\cdots]] e^{\hat B}
\end{eqnarray}
and
\begin{eqnarray}
  &&{d\over dt}e^{\hat B(t)}=\biggl( {d \hat B(t)\over t}+
  \sum_{n=1}^\infty{1\over (n+1)!}
  [\hat B(t),[\cdots,[\hat B(t),{d \hat B(t)\over dt}]\cdots]]
  \biggr) e^{\hat B(t)}
\end{eqnarray}
are satisfied. Using these relation we have
\begin{eqnarray}
  &&\hnabla_\alpha \Psi^A(x,\pm y)=e^{\pm y^\nu\hnabla_\nu}
  \hnabla_\alpha \psi^A(x)-\hat H_\alpha(x,\pm y)\Psi^A(x,\pm y)
\label{B2}
\end{eqnarray}
and
\begin{eqnarray}
  &&{\partial\over \partial y^\alpha}\Psi^A(x,\pm y)=
  \pm\hnabla_\alpha \Psi^A(x,\pm y)\pm\hat G_\alpha(x,\pm y)\Psi^A(x,\pm y),
\label{B8}
\end{eqnarray}
where we defined
\begin{eqnarray}
  &&\hat H_\alpha(x,\pm y)=\sum_{n=1}^\infty{(\pm1)^n\over n!} 
  y^{\nu_1}\cdots y^{\nu_n} [\hnabla_{\nu_1},[\cdots,[\hnabla_{\nu_n},\hnabla_\alpha]\cdots]]
\end{eqnarray}
and
\begin{eqnarray}
  &&\hat G_\alpha(x,\pm y)=\sum_{n=1}^\infty{(\pm1)^n\over (n+1)!} 
  y^{\nu_1}\cdots y^{\nu_n} [\hnabla_{\nu_1},[\cdots,[\hnabla_{\nu_n},\hnabla_\alpha]\cdots]],
\end{eqnarray}
respectively. We also have
\begin{eqnarray}
  &&e^{\pm y^\nu\hnabla_\nu} \sum_B \Bigl(M^{AB}(x)
  +\hbar\gamma^\mu J_\mu^{AB}(x)\Bigr)\psi^B(x)
\nonumber
\\
  &&\hspace{1cm}
  =\sum_B\Bigl(M^{AB}(x)+\hbar\gamma^\mu J_\mu^{AB}(x)\Bigr)\Psi^B(x,\pm y)
  +\sum_B\hat L^{AB}(x,\pm y)\Psi^B(x,\pm y),
\label{B5}
\end{eqnarray}
with defining
\begin{eqnarray}
  &&\hat L^{AB}(x,\pm y)=\sum_{n=1}^\infty{(\pm1)^n\over n!}
  y^{\nu_1}\cdots y^{\nu_n} \nabla_{\nu_1}\cdots\nabla_{\nu_n}
  \Bigl(M^{AB}(x)+\hbar\gamma^\mu J_\mu^{AB}(x)\Bigr).
\end{eqnarray}

Next we consider $D_\alpha N^{AB}(x,p)$, where $D_\alpha$
is defined by (\ref{defD}).
Using eqs.(\ref{B2}) and (\ref{B8}), we have the expression:
\begin{eqnarray}
  \hatD_\alpha N^{AB}(x,p)&=&{2i\over \hbar} p_\alpha N^{AB}(x,p)-
  {\sqrt{-g(x)}\over (\pi\hbar)^4}\int d^4y e^{-2iy^\beta p_\beta/\hbar}
\nonumber
\\
  &&\times  \Bigl\langle 2(e^{-y^\nu\hnabla_\nu}\hnabla_\alpha \psi^A(x))
  \bar\Psi^B(x,y)-G_\alpha^{AB}(x,y)\Bigr\rangle,
\label{kea}
\end{eqnarray}
where 
\begin{eqnarray}
  G_\alpha^{AB}(x,y)&=&
  \Psi^A(x,-y)\Bigl(\hat G_\alpha(x,y)\bar \Psi^B(x,y)\Bigr)
  -\Bigl(\hat G_\alpha(x,-y)\Psi^A(x,-y)\Bigr)\bar\Psi^B(x,y)
\nonumber
\\
  &&+2\Bigl(\hat H_\alpha(x,-y)\Psi^A(x,-y)\Bigr)\bar\Psi^B(x,y).
\label{GAB}
\end{eqnarray}
Projecting $i\hbar\gamma^\alpha(x)/2$ on the both side of (\ref{kea}), 
we have 
\begin{eqnarray}
  &&\gamma^\alpha(x) \biggl(p_\alpha+{i\hbar\over 2}\hatD_\alpha \biggr)
 N^{AB}(x,p)
  =-{i\hbar \over 2}{\sqrt{-g(x)}\over (\pi\hbar)^4}\int d^4y e^{-2iy^\beta p_\beta/\hbar}
\nonumber
\\
  &&\hspace{4cm}
  \times\gamma^\alpha(x)\Bigl\langle 2(e^{-y^\nu\hnabla_\nu}\hnabla_\alpha \psi^A(x))
  \bar\Psi^B(x,y)-G_\alpha^{AB}(x,y)\Bigr\rangle.
\label{keb}
\end{eqnarray}
Using the Dirac equation (\ref{DiracE}) 
and 
\begin{equation}
  \nabla_\alpha\gamma^\beta(x)=0,
\end{equation}
we have
\begin{eqnarray}
  &&\gamma^\alpha(x) \biggl(p_\alpha+{i\hbar\over 2}\hatD_\alpha \biggr)N^{AB}(x,p)
  =\sum_C \calM^{AC}\Bigl(x,
  {-i\hbar\over 2}{\partial\over \partial p}\Bigr)N^{CB}(x,p)
\nonumber
\\
  &&\hspace{4cm}
  +{i\hbar \over 2}{\sqrt{-g(x)}\over (\pi\hbar)^4}\int d^4y e^{-2iy^\beta p_\beta/\hbar}
  \gamma^\alpha(x)\Bigl\langle G_\alpha^{AB}(x,y)\Bigr\rangle,
\label{ked}
\end{eqnarray}
where 
\begin{eqnarray}
  \calM^{AB}\Bigl(x,
  {-i\hbar\over 2}{\partial\over \partial p}\Bigr)&=&
  M^{AB}(x)+\hbar\gamma^\mu J_\mu^{AB}
\nonumber
\\
  &&\hspace{0.2cm}+\sum_{n=1}^\infty
  {1\over n!}
  \Bigl({-i\hbar\over 2}\Bigr)^n
  \nabla_{\nu_1}\cdots\nabla_{\nu_n}
  \Bigl(M^{AB}(x)+\hbar\gamma^\mu J_\mu^{AB}\Bigr)
  {\partial\over \partial p_{\nu_1}}
  \cdots 
  {\partial\over \partial p_{\nu_n}}.
\nonumber
\\
\end{eqnarray}
Then we focus on the last term of the right hand side of (\ref{ked}).
Using $C_k^n$ to denote the binomial coefficient, one can derive the 
following recursion formula:
\begin{eqnarray}
  G_\alpha^{A(n)}=-A_\alpha^{(n-1)}\Psi^A(x,-y)
  +R_\alpha^{(n)\nu}{\partial\over \partial y^\nu}\Psi^A(x,-y)
  -\sum_{k=1}^{n-2}C_k^{n-2}R_\alpha^{(n-k)\nu}G_\nu^{A(k)},
\label{recursionA}
\end{eqnarray}
where we defined
\begin{eqnarray}
  &&G_\alpha^{A(n+1)}=y^{\nu_1}\cdots y^{\nu_n}
  [\hnabla_{\nu_1},[\cdots,[\hnabla_{\nu_n},\hnabla_\alpha]\cdots]]\Psi^{A}(x,-y) 
  \hspace{1cm}(n\geq 1),
\nonumber
\\
  &&G_\alpha^{A(1)}=\hnabla_\alpha\Psi^{A}(x,-y), 
\end{eqnarray}
\begin{eqnarray}
  &&R_\alpha^{(n)\beta}=y^{\nu_1}\cdots y^{\nu_n}
  R^{\beta}_{\nu_1\alpha\nu_2;\nu_3\cdots\nu_n}
  \hspace{1cm}(n\geq 3),
\nonumber
\\
  &&R_\alpha^{(2)\beta}=y^{\nu_1}y^{\nu_2}R^{\beta}_{\nu_1\alpha\nu_2},
\end{eqnarray}
\begin{eqnarray}
  &&A_\alpha^{(n)}=y^{\nu_1}\cdots y^{\nu_n}A_{\alpha\nu_1;\nu_2\cdots\nu_n}
  \hspace{1cm}(n\geq 2),
\nonumber
\\
  &&A_\alpha^{(1)}=y^{\nu_1}A_{\alpha\nu_1},
\nonumber
\end{eqnarray}
with
\begin{eqnarray}
  A_{\alpha\beta}={1\over 4}R_{\alpha\beta\mu\nu}\gamma^{\mu}\gamma^{\nu},
\label{Aab}
\end{eqnarray}
which satisfies the commutator relation 
$
  [\nabla_\alpha,\nabla_\beta]\psi^A(x)=A_{\alpha\beta}\psi^A(x).
$
%
In a similar way, we also have the following recursion formula
\begin{eqnarray}
  \bar G_\alpha^{B(n)}=\bar\Psi^B(x,y)A_\alpha^{(n-1)}
  +R_\alpha^{(n)\nu}{\partial\over \partial y^\nu}\bar\Psi^B(x,y)
  -\sum_{k=1}^{n-2}C_k^{n-2}R_\alpha^{(n-k)\nu}\bar G_\nu^{B(k)},
\label{recursionB}
\end{eqnarray}
with
\begin{eqnarray}
  &&\bar G_\alpha^{B(n+1)}=y^{\nu_1}\cdots y^{\nu_n}
  [\hnabla_{\nu_1},[\cdots,[\hnabla_{\nu_n},\hnabla_\alpha]\cdots]]\bar \Psi^{B}(x,y) 
  \hspace{1cm}(n\geq 1)
\nonumber
\\
  &&\bar G_\alpha^{B(1)}=\hnabla_\alpha\bar \Psi^{B}(x,y).
\end{eqnarray}

A Repeated use of the recursion formula (\ref{recursionA}) or (\ref{recursionB})
allows one to rewrite $\hat G_\alpha(x,-y)\Psi^A(x,-y)$
and $\hat H_\alpha(x,-y)\Psi^A(x,-y)$ or $\hat G_\alpha(x,y)\bar\Psi^B(x,y)$
in equation (\ref{GAB}), 
in terms of $\partial \Psi^A(x,-y)/\partial y^\nu$, $\hnabla_\nu \Psi^A(x,-y)$
and $\Psi^A(x,-y)$, eliminating the derivative terms higher than 
the second derivatives. 
Then the formal expression for $G_\alpha^{AB}$ is written in the form, 
\begin{eqnarray}
  G_\alpha^{AB}(x,y)&=&
  \biggl[\sum_{k=2}^\infty K^{(1)\beta}_{(k)\alpha}R^{(k)\nu}_\beta\hnabla_\nu 
  -\sum_{k=2}^\infty K^{(2)\beta}_{(k)\alpha}R^{(k)\nu}_\beta{\partial \over
  \partial y^\nu}\biggr] \Bigl(\Psi^A(x,-y)\bar \Psi^B(x,y)\Bigr)
\nonumber
\\
  &&+\sum_{k=2}^\infty K^{(3)\beta}_{(k)\alpha}A^{(k-1)}_\beta
  \Bigl(\Psi^A(x,-y)\bar \Psi^B(x,y)\Bigr)
\nonumber
\\
  &&+\Bigl(\Psi^A(x,-y)\bar \Psi^B(x,y)\Bigr)
  \sum_{k=2}^\infty K^{(4)\beta}_{(k)\alpha}A^{(k-1)}_\beta,
\label{GABEX}
\end{eqnarray}
where 
\begin{eqnarray}
  K^{(i)\beta}_{(k)\alpha}=\delta^\beta_\alpha C^{(i)}_{(k)} +
  \sum_{n=1}^\infty\sum_{k_1=2}^\infty\cdots\sum_{k_n=2}^\infty
  C^{(i)}_{(k)k_1\dots k_n}R^{(k_n)\beta_n}_\alpha\cdots
  R^{(k_1)\beta}_{\beta_2},
\end{eqnarray}
and $C^{(i)}_{(k)}$ and $C^{(i)}_{(k)k_1\cdots k_n}$ are numerical 
coefficients, which can explicitly be found using generating 
functional method \cite{FonarevJMP}. We omit the general expression, 
however, the lowest order coefficients are, for example,
\begin{eqnarray}
  &&C_{(k)}^{(1)}={(-1)^k\over k!}+{1-(-1)^k\over (k+1)!},
\\
  &&C_{(k)}^{(2)}={(-1)^k\over k!}-{1+(-1)^k\over (k+1)!},
\\
  &&C_{(k)}^{(3)}={(-1)^k(2k-1)\over k!},
\\
  &&C_{(k)}^{(4)}={1\over k!}.
\end{eqnarray}
Using (\ref{GABEX}), we have
\begin{eqnarray}
  &&  {i\hbar \over 2}{\sqrt{-g(x)}\over (\pi\hbar)^4}\int d^4y 
  e^{-2iy^\beta p_\beta/\hbar}
  \Bigl\langle G_\alpha^{AB}(x,y)\Bigr\rangle=
\nonumber
\\
  &&\hspace{1cm}
  -\sum_{k=2}^\infty\biggl[
  \biggl(\calK^{(2)\beta}_{(k)\alpha}\calR^{(k)\nu}_\beta p_\nu
  +{i\hbar\over 2}\calK^{(1)\beta}_{(k)\alpha}\calR^{(k)\nu}_\beta \hatD_\nu
  \biggr)N^{AB}(x,p)
\nonumber
\\
  &&\hspace{1cm}
  +{i\hbar\over 2}\calK^{(3)\beta}_{(k)\alpha}\calA^{(k-1)}_\beta N^{AB}(x,p)
  +{i\hbar\over 2}N^{AB}(x,p)\calK^{(4)\beta}_{(k)\alpha}\calA^{(k-1)}_\beta
  \biggr],
\label{keh}
\end{eqnarray}
where
\begin{eqnarray}
  &&\calK^{(i)\beta}_{(k)\alpha}=\delta^\beta_\alpha C^{(i)}_{(k)} +
  \sum_{n=1}^\infty\sum_{k_1=2}^\infty\cdots\sum_{k_n=2}^\infty
  C^{(i)}_{(k)k_1\dots k_n}\calR^{(k_n)\beta_n}_\alpha\cdots
  \calR^{(k_1)\beta}_{\beta_2},
\\
  &&\calR^{(k)\beta}_\alpha=\biggl({i\hbar\over2}\biggr)^k
  R^\beta_{\nu_1\alpha\nu_2;\nu_3\cdots\nu_k}
  {\partial\over \partial p_{\nu_1}}\cdots
  {\partial\over \partial p_{\nu_k}},
\\
  &&\calA^{(k)}_\alpha=\biggl({i\hbar\over2}\biggr)^k
  A_{\alpha\nu_1;\nu_2\cdots\nu_k}
  {\partial\over \partial p_{\nu_1}}\cdots
  {\partial\over \partial p_{\nu_k}}.
\end{eqnarray}
Finally we find that equation $N^{AB}(x,p)$ follows
\begin{eqnarray}
  &&\gamma^\alpha(x) \biggl(p_\alpha+{i\hbar\over 2}\hatD_\alpha \biggr)N^{AB}(x,p)
   -\sum_C\calM^{AC}\Bigl(x,{-i\hbar\over 2}{\partial\over \partial p}\Bigr)N^{CB}(x,p)
\nonumber
\\
  &&\hspace{1cm}=  -\gamma^\alpha(x)\sum_{k=2}^\infty\biggl[
  \biggl(\calK^{(2)\beta}_{(k)\alpha}\calR^{(k)\nu}_\beta p_\nu
  +{i\hbar\over 2}\calK^{(1)\beta}_{(k)\alpha}\calR^{(k)\nu}_\beta \hatD_\nu
  \biggr)N^{AB}(x,p)
\nonumber
\\
  &&\hspace{1cm}
  +{i\hbar\over 2}\calK^{(3)\beta}_{(k)\alpha}\calA^{(k-1)}_\beta N^{AB}(x,p)
  +{i\hbar\over 2}N^{AB}(x,p)\calK^{(4)\beta}_{(k)\alpha}\calA^{(k-1)}_\beta
  \biggr].
\label{kei}
\end{eqnarray}

\end{appendix}

\end{document}